\documentclass[aps,prl,twocolumn,showpacs,floatfix,footinbib,groupedaddress,superscriptaddress]{revtex4}
\usepackage{amsmath}
\usepackage{amssymb,graphics,graphicx,times,bm,bbm,multirow}

\begin{document}

\title{Symmetry-enhanced supertransfer of delocalized quantum states}

\author{Seth Lloyd}
\affiliation{Department of Mechanical Engineering, Massachusetts
Institute of Technology, 77 Massachusetts Avenue, Cambridge MA
02139} \affiliation{Center for Extreme Quantum Information Theory,
Research Laboratory of Electronics, Massachusetts Institute of
Technology, Cambridge, MA 02139}
\author{Masoud Mohseni}
\affiliation{Center for Extreme Quantum Information Theory, Research
Laboratory of Electronics, Massachusetts Institute of Technology,
Cambridge, MA 02139} \affiliation{Center for excitonics, Research
Laboratory of Electronics, Massachusetts Institute of Technology,
Cambridge, MA 02139}

\begin{abstract}

Coherent hopping of excitation rely on quantum coherence over
physically extended states. In this work, we consider simple models
to examine the effect of symmetries of delocalized multi-excitation
states on the dynamical timescales, including hopping rates,
radiative decay, and environmental interactions. While the
decoherence (pure dephasing) rate of an extended state over $N$
sites is comparable to that of a non-extended state, superradiance
leads to a factor of N enhancement in decay and absorption rates. In
addition to superradiance, we illustrate how the multi-excitonic
states exhibit `supertransfer' in the far-field regime: hopping from
a symmetrized state over N sites to a symmetrized state over M sites
at a rate proportional to MN. We argue that such symmetries could
play an operational role in physical systems based on the
competition between symmetry-enhanced interactions and localized
inhomogeneities and environmental interactions that destroy
symmetry. As an example, we propose that supertransfer and coherent
hopping play a role in recent observations of anomolously long
diffusion lengths in nano-engineered assembly of light-harvesting
complexes.

\end{abstract}

\maketitle

Recent works on quantum coherence of photosynthesis has revealed a
rich tapestry of coherent and incoherent interactions in excitonic
hopping processes
\cite{Engel07,Lee07,Mohseni08,Plenio08-1,Mercer09,Scholes09-1,Scholes09-2,Jang08,Ishizaki09,Escalante10}.
Natural selection may well have tuned the tradeoffs between coherent
hopping, decoherence, dissipation, and the non-Markovian nature of
the phonon environment, to arrive at robust and highly efficient
energy transfer methods \cite{Mohseni08,Ishizaki09}. As is typical
of biological systems, the highly-evolved final product of evolution
exhibits a complexity which reflects the different types of
apparatus required to take advantage of the rich range of dynamical
effects at the microscale. While efficient from an evolutionary
standpoint, this complexity can mask the essential simplicity of the
underlying the quantum effects that allow energy transfer to take
place in the first place.

One of the key features of excitonic energy transfer in
photosynthesis is the extended, delocalized nature of the states
involved in the hopping process.  This extended form of the states
is enforced from the very moment of photon absorption: because the
wavelength of light is large compared with the atomic scale, a very
large numbers of electrons participate coherently in the initial
absorption process, and the resulting initial excitonic state
extends over a large number of atoms or molecules. This paper looks
at the dynamics and environmental interactions of states that extend
over $N$ atoms, and uses simple quantum models to establish some
general features of such interactions.

First of all, states that interact symmetrically with a collection
of bosons (phonons or photons or both), exhibit the normal $N$-fold
superradiant enhancement of emission and absorption in both the
resonant and non-resonant regimes
\cite{Dicke54,RehlerEberly71,Fidder91,Zhao99,Palacios02,Jin03}.
Superradiance, of course, follows directly from the symmetrized
nature of the atom-boson interaction.  Interactions that break that
symmetry can lead to non-radiant excited states, multiple-site
analogues of the non-radiant two-site antisymmetrized singlet state.
By contrast, environmental interactions such as decoherence that do
not rely on energy exchange exhibit no `super' enhancement of the
decoherence rate of single-exciton states due to symmetry.

We employ a second-quantized spin-boson notation that allows the
treatment of multiple excitonic states. For multiple excitonic
states, the decoherence rate can either be enhanced or decreased,
depending on the nature of the correlations between the multiple
excitons.  Symmetrized states of $n$ excitons spread amongst $N$
sites exhibit the usual
superradiant factor of $(N-n+1) n$ in their emission rate.
Antisymmetrized states of $n$ excitons can have their emission rate
largely suppressed, thereby providing a potential mechanism
for exciton preservation in multi-site systems.

Decoherence or pure dephasing has a different dependency on exciton
number: symmetrized states of $n$ excitons coupled symmetrically to
a common environment exhibit a decoherence rate that goes as $n$
times the single exciton rate.  In contrast, the decoherence rate of
an uncorrelated $n$-exciton state, or a symmetric state coupled to
an asymmetric environment, typically goes as $\sqrt n$ times the
single exciton rate.  As in the case of emission, states in special
antisymmetrized decoherence-free subspaces can have their
decoherence rates largely suppressed.

The features of delocalized states interacting symmetrically with
their environment described so far are well-known. In particular,
the effects of inhomogeneous broadening (static disorder) and
phonon-bath coupling on superradiance relaxation for molecular
aggregates have been studied in detail
\cite{Fidder91,Zhao99,Palacios02}. Less familiar is the phenomenon
of cooperative excitation energy transfer or supertransfer
\cite{Strek77,Scholes02}: consider an extended excitonic state over
$N$ sites that is hopping to an extended state over $M$ sites. If
the hopping interaction possesses the proper symmetries, then we
show below that the overall hopping rate is proportional to $MN$,
and enhancement of $N$ over the hopping rate of a localized state
hopping to one of $M$ sites. The supertransfer enhancement follows
from the same symmetry considerations as superradiance, but it is
essentially a radiationless relaxation. Under certain conditions
where a molecular aggregate can coherently donate an excitation and
an acceptor molecular aggregate can coherently receive it, then the
dipole approximation and first-order perturbation for electronic
coupling imply that we might observe supertransfer of excitation
energy. Supertransfer also appears when the time-averaged site-site
couplings $\gamma_{ij}$ between each of the $N$ sites and each of
the $M$ sites are largely similar. Cooperative excitation energy
transfer can also occur in disordered chromophoric systems and
light-harvesting complexes \cite{Scholes02}.

Symmetric enhancements of dynamical timescales extend to
environmental interactions. An interaction between excitons at $N$
sites and a set of bosonic modes (phonons, photons) can be
decomposed into fully symmetrized interactions between the symmetric
states over those sites and symmetrized bosonic modes, and into
other interactions with different symmetries. The fully symmetrized
interactions participate in the $N$-fold superradiant enhancement of
interaction rates. Accordingly, these interactions participate more
strongly in, e.g., symmetrized hopping interactions. Meanwhile, the
other interactions tend to destroy the symmetry of the extended
states and reduce hopping.

A full treatment of the various symmetries of interactions between
spins, atoms, excitons, and bosonic environment modes would require
a general treatment in terms of representations of the symmetric
group using Young diagrams and tableaux. As the purpose of this
paper is simply to examine the the way in which superradiant
enhancements `spill over' into hopping and environmental
interactions, we content ourselves here with a treatment of fully
symmetrized states and leave the more general treatment for
elsewhere \cite{Mohseni10}.

\section{Supertransfer in spin-boson model}

We start by reviewing the ordinary picture of superradiance in the
resonant interaction between $N$ two-level atoms and a mode of the
electromagnetic field.  The Hamiltonian for this system is:
$$H = \hbar \big( \omega a^\dagger a - \omega/2 \sum_{j=1}^N
\sigma_z^j + \gamma \sum_j \sigma_x^j(a+a^\dagger) \big).\eqno(1)$$
For simplicity we have assumed that the oscillatory field is
polarized along the $x$-axis.  The ground state of each atom is
$|0\rangle = |\uparrow\rangle$, and the excited state is $|1\rangle
= |\downarrow\rangle$. The fully symmetrized state with $n$ excited
atoms is
$$|n\rangle \equiv {1\over \sqrt{N\!}} \sum_{\pi \in {\cal S}_N}
| \pi(00\ldots 0 11\ldots 1) \rangle. \eqno(2)$$ Here $\pi$ is a
permutation in the symmetric group over $N$ elements, ${\cal S}_N$,
and there are $n$ $1$'s and $N-n$ $0$'s in the state. The states
$|n\rangle$ are those that are obtained from the state $|N\rangle =
|11\ldots 1\rangle$ by radiant decay, or from the state $|0\rangle =
|00\ldots 0\rangle$ by stimulated absorption.

As Dicke pointed out, superradiance arises from the symmetrized
nature of the states and their interaction with the field
\cite{Dicke54}. If we look at the decay rate of a single atom
coupled to the field in its vacuum state, we find that its amplitude
to first order in perturbation theory goes as $\gamma$, and its
probability goes as $\gamma^2$.  By contrast, the decay amplitude of
the symmetrized state $|n\rangle$ goes as
\begin{align}
\langle m= 1| \langle n-1| \gamma \sum_j \sigma^j_x (a^\dagger + a)
|n\rangle |m=0\rangle \cr = \sqrt{ n(N-n+1)} \gamma, \tag{3}
\end{align}
where $m$ labels the photon number in the mode. The decay
probability goes as $n(N-n+1) \gamma^2$.   Comparing the decay rate
of the one-excitation symmetrized state $|n=1\rangle$ with that of
the single-atom decay rate, we see that the symmetrized state decays
$N$ times as fast.  By contrast, the stimulated emission rate for
the symmetrized state is the same as the incoherent stimulated
emission rate.

Symmetry can enhance more than the spontaneous emission rate. Let's
turn to hopping and look what happens when an excitonic state that
is symmetrized over $N$ sites hops to an excitonic state that is
symmetrized over $M$ sites via a symmetric coupling \cite{Strek77}.
In this idealized case, the symmetrized hopping Hamiltonian is
$$H = \hbar\big( - {\omega_A \over 2} \sum_{j=1}^N \sigma_z^j
- {\omega_B \over 2} \sum_{k=1}^M \sigma_z^k
 + \gamma \sum_{j=1, k=1}^{N,M} \sigma_+^j \sigma_-^k
+\sigma_-^j \sigma_+^k \big).\eqno(4)$$ Here, $j$ labels the $A$
sites and $k$ labels the $B$ sites. Performing the same calculation
as in superradiance, but taking into account the symmetrized nature
of both the $A$ and $B$ states, shows that the rate of hopping of an
excitation from the symmetrized state $|n\rangle |m\rangle$ to the
symmetrized state $|n-1\rangle|m+1\rangle$ goes as $\gamma^2
n(N-n+1) (m+1)(M-m)$.  Taking into account transfer from $B$ to $A$
as well as from $A$ to $B$, we find that the overall transfer rate
from $A$ to $B$, starting in the state $|n\rangle |m\rangle$, goes
as
$$\gamma^2 \big(
n(N-n+1) (m+1)(M-m) - (n+1)(N-n) m(M-m+1)\big),\eqno(5)$$
where the first
term represents the rate of an excitation hopping from $A$ to
$B$ and the second term represents the rate of hopping from $B$
to $A$.  For a single symmetrized excitation hopping
from the $A$ states to the $B$ states we see that the rate of
transition from $|n=1\rangle |m=0\rangle$ to $|n=0\rangle
|m=1\rangle$ goes as $\gamma^2 NM$: that is, the hopping rate
between symmetrized single excitation states goes as $NM$ times the
hopping rate of an excitation between one of the $A$ sites to one of
the $B$ sites.  For higher numbers of excitations, we see
that the quartic terms in equation (5) cancel out, leaving
only cubic terms.  For example, when there are $O(N)$ excitations in $A$,
and $O(M)<O(N)$ excitations in $B$, the transfer rate
from $A$ to $B$ goes as $O(N^2M)$.

Similar supertransfer due to partial symmetry may be relevant to
various hopping processes in photosynthetic light-harvesting (LH)
complexes \cite{Fidder91,Zhao99,Palacios02}, within the context of
generalized F\"orster theory
\cite{Scholes00,Scholes01,Scholes02,JangSilbey04}. In the typical
treatment of Frenkel exciton in these systems, the transfer rate is
calculated from the transition probability of an excitation hopping
from one molecule to another using F\"orster Resonance Energy
Transfer (FRET) based on dipole-dipole interaction of individual
molecules and perturbation theory (Fermi's golden rule). However,
due to strong interactions of a group of molecules the excitation
can become highly delocalized. Thus, one can introduce a huge
(effective) dipole moment associated with each group leading to an
enhanced oscillator strength. Consequently, the rate of exciton
dynamics can be calculated from these effective very large
dipole-dipole interactions even in the far-field within the dipole
approximation. A closely packed group of N molecules under certain
symmetry can collectively accept or donate an excitation with a rate
which is almost N times faster than each individual molecule.

Consider, for example, the rate of hopping of a single exciton from
a ring containing $N$ chromophores to a ring containing $M$
chromophores. If the chromophores are spread evenly along each ring,
then the lowest energy states single-exciton states in each ring are
in fact the fully symmetric $|n=1\rangle$ and $|m=1\rangle$ states.
If the two rings are distant from each other and coupled, e.g., via
dipolar F\"orster forces, then the couplings of the chromophores
between rings are approximately symmetric. Accordingly, the rate of
hopping from the $A$ ring to the $B$ ring is $NM$ times the rate of
the rate of an exciton hopping from a chromophore on the $A$ ring to
a chromophore on the $B$ ring. The hopping enhancement arises
because hopping between symmetrized states is essentially a kind of
superradiance where the $A$ states `emit' their excitonic energy
into the $B$ states. Circular symmetry of the above nature exists in
LHI and LHII of purple bacteria \cite{Damjanovi97}.

\section{Symmetry properties of delocalized states under environmental
interactions} There are many other effects in real quantum systems
including diagonal and off-diagonal static disorders, couplings of
excitonic states within a single ring, and phonon-bath couplings. If
$\omega_A \neq \omega_B$, then the excitonic transfer must be
accompanied by energy transfer to/from the environment. Now, just as
the effects of symmetry translated directly over from superradiance
to supertransfer, we will show that symmetry can similarly enhance
the effects of environmental interactions. Let's include intra-ring
couplings and add bosonic environments to the $A$ and $B$ states, so
that our overall Hamiltonian is
\begin{align}
H =  \hbar\bigg( &- {\omega_A \over 2} \sum_{j=1}^N \sigma_z^j
+\sum_{\ell} \omega_{A\ell} a_\ell^\dagger a_\ell + \sum_{j\ell}
\Gamma_{j\ell} H_{j\ell}\cr &- {\omega_B \over 2} \sum_{k=1}^M
\sigma_z^k +\sum_{\ell'} \omega_{B\ell'} a_{\ell'}^\dagger a_{\ell'}
+ \sum_{k\ell'} \Gamma_{j\ell'} H_{j\ell'}\cr &+ \gamma \sum_{j=1,
k=1}^{N,M} \sigma_+^j \sigma_-^k +\sigma_-^j \sigma_+^k\cr &+
\sum_{j,j'=1}^N \gamma_{jj'} (\sigma_+^j \sigma_-^{j'} +\sigma_-^j
\sigma_+^{j'})\cr &+ \sum_{k,k'=1}^M \gamma_{kk'} (\sigma_+^k
\sigma_-^{k'} +\sigma_-^k \sigma_+^{k'})
 \bigg). \cr
\tag{6}
\end{align}
Here $H_{j \ell}$ is some suitable interaction between the $j$'th
spin and the $\ell$'th mode, e.g., $\gamma_{j\ell}(
a^\dagger_\ell\sigma^j_- + a_\ell \sigma^j_+)$.

The $\gamma$ term represents the symmetric inter-ring coupling, and
the $\gamma_{jj'}$, $\gamma_{kk'}$ terms represent the intra-ring
couplings within the $A$ and $B$ rings respectively. Note that even
with intra-ring hopping couplings, because of the symmetric
arrangement of chromophores in the ring, the fully symmetric single
exciton state within each ring is the ground state of the
single-exciton sector.

The key feature of this Hamiltonian (and of similar
multiple-spin/multiple boson models) is that the environmental
interaction can contain a significant component that is coupled
directly to the symmetrized states $|n\rangle$.  This is true even
if the general interaction takes the form of interactions between
local sites and local modes as above.  The insight here is that,
because the interaction is linear in the $a_\ell, a_\ell^\dagger$
and the Pauli matrices, we are always free to perform a Bogoliubov
transformation on the modes to identify modes corresponding to
delocalized, symmetrized bosonic excitations.  In the case of
chromophoric ring, as above, these excitations are simply the global
symmetrized vibrational modes of the ring itself. That is, if each
site is coupled to a localized phonon mode with frequency $\omega$,
then we are free to define a delocalized, symmetric phonon mode with
frequency $\omega$: the symmetrized states of the ring are then
coupled to this symmetrized phonon state with the usual superradiant
factor of $N$.

Now, just as in superradiance, the amplitude in first order
perturbation theory for the destruction of a symmetrized excitation
of the spins and the creation of a symmetrized excitation of the
bosonic modes is proportional to $\sqrt{n(N-n+1)}$.  Moreover, in
the overall Hamiltonian (6), we can extract out the symmetrized
sectors of spin states and bosonic states in $A$ and $B$
respectively. Transitions between these states, including those that
involve emission of energy into the symmetrized phonon modes, all
involve superradiant enhancements.  As the case of hopping shows,
when an $N$-site symmetrized state exchanges energy via a symmetric
interaction with an $M$-site state, the enhancement of the
interaction rate goes as the product $n(N-n+1) (m+1)(M-m)$, where
$n,m$ are the excitation number of the symmetrized states, including
now the states of the symmetrized bosonic modes, and the net
transfer of energy goes as equation (5).  While the
case of excitonic hopping is frequently restricted physically
to the interactions of a few excitons at a time, the symmetrized
modes of the bosonic bath can readily contain a large number of
phonons, so that the extra factors of $n,m$ can really `kick in'
and enhance the transfer of energy from excitonic
states to the bath, and back to excitonic states again.
As equation (5) shows, the rate of transfer grows as a cubic function
of the populations and site numbers.

We can now formally decompose the Hamiltonian of equation (6)
into cooperative and `normal' sectors.  The cooperative sector
consists of the symmetrized states over the $N$ sites
of the $A$ sector and the $M$ sites of the $B$ sector,
together with the symmetrized states of the bosonic modes in each
sector.  Let $P_C$ be the projection operator onto the
subspace of the system-environment Hilbert space that is spanned by
the cooperative states.  Similarly, let $P_N = 1-P_C$ be the
projector onto the `normal,' or non-cooperative subspace. We can
then decompose our general Hamiltonian, equation (6), into
cooperative sectors (C) and a `normal' sector (N), together with a
couplings between these sectors: $H= H_C + H_N + H_{CN}$, where $H_C
= P_C H P_C$ is the Hamiltonian confined to the cooperative
subspace, $H_N = P_N H P_N$, is the Hamiltonian confined to the
normal subspace, and $H_{CN} = P_C H P_N + P_N H P_C$ is the part of
the Hamiltonian that couples the cooperative to the normal sector.

In the single-exciton sector, $H_C$ just represents two
two-level systems, each coupled to its own environment, but with an
enhancement of $N M$ in hopping strength of the exciton
from $A$ to $B$, and with an enhancement of
of $N (m+1)(N-m)$ in the interaction strengths
for exchange of energy between the $A$-exciton and the
the $m$ bosons in the cooperative mode of $A$'s environment
(similarly for $B$).  In the multiple-exciton sector,
$H_C$ is a Hamiltonian that couples
two nonlinear harmonic oscillators, where the nonlinearity
arises from excitonic interactions; $H_C$ also contains these
oscillators' interactions with their cooperative environments.

The two essential features of the decomposition into cooperative and
normal sections are as follows. First, the cooperative Hilbert space
has a drastically reduced dimension compared with the full Hilbert
space. From the point of view of transport efficiency, this reduced
dimension is useful because it prevents the hopping exciton from
becoming `lost in Hilbert space': the exciton can inhabit either the
$A$ oscillator, or the $B$ oscillator, or it can be transferred to
the cooperative environment.  The cooperative Hilbert space is too
simple, however, to admit such phenomena as localization.  In other
words, the cooperative Hilbert space is simply too small to get lost
in.   From the perspective of a scientist investigating the behavior
of excitonic hopping, the small size of the cooperative Hilbert
space has the advantage that simulating behavior of the cooperative
sector is relatively simple compared with a full many-body
treatment.

The second key feature of the full Hamiltonian
$H_C + H_N + H_{CN}$ is the relative strength of
the dynamics in the different sectors.  In the cooperative
sector, interactions cooperate coherently, leading to
an enhancement of $N,M$ for the interactions of $A$ and $B$
with their respective environments, and an enhancement of
$MN$ for the interaction strength.  By contrast,
the interactions between the cooperative and normal
sectors involve mixed symmetries which add incoherently.
The terms $H_{CN}$ couple the fully
symmetrized sector of Hilbert space to sectors with differing
symmetry.  These parts of the Hamiltonian induce `leakage' from the
symmetric sector.  Because the leakage rates do not receive any
superradiant enhancement, they can in principle be modelled as a
perturbation to the cooperative dynamics.
To fully explore the implications of the existence
of the cooperative sector, we should construct a master equation
using the cooperative sector as `system' and the non-cooperative
sector as `environment.'  Such a symmetry-based master
equation approach will be explored in future work.

Our overall picture,
then is as follows: we have a cooperative sector where quantum
interference and symmetrization induce enhanced rates
of hopping and energy exchange with a symmetrized environment.
The cooperative sector has relatively few degrees of freedom.
The `non-cooperative,' i.e., ordinary, sector has many
more degrees of freedom, but the coupling rates from the
cooperative to uncooperative sector can be considerably
smaller (e.g, two orders of magnitude smaller if $A$ and $B$
each have a dozen or so sites) than the coupling rates
within the cooperative sector.
Once a state departs from the cooperative sector, however,
it is unlikely to return unless the cooperative sector possesses
an intrinsically lower energy than states in the non-cooperative
sector, in which case relaxation can drive the system back
into the cooperative sector.  Resymmetrization by relaxation
occurs, for example, in the common case where the lowest energy
single exciton state in a ring is the state that is a symmetric
superposition of exciton located at each chromophore.  (In quantum
information, such a state is known as a $W$ state.)

Resymmetrization via relaxation of excitonic states within
a ring has the potential to increase the hopping rate by
the following mechanism.  In situations where the spatial
extent of the ring is not small compared with the distance
between rings, the couplings of excitons between rings will
no longer be symmetric: excitons at the closer edge will couple
more strongly than excitons at the further edge.  Consequently,
the symmetrized state of one ring may have a significant
coupling to an asymmetric state of the second ring.
As long as these asymmetric states relax to the ground state
of the second ring, then such coupling will tend to
increase the transfer rate.

A second way that the transfer rate can be enhanced is by coupling
of excited states of ring $A$ with excited states of the same
symmetry class of ring $B$. The $N$-dimensional Hilbert space of
single-excitonic states of $A$ decomposes under the symmetric group
into the direct sum of the one-dimensional fully-symmetric space --
the fully-symmetric state described above -- plus the $N-1$
dimensional antisymmetric subspace. If the ambient temperature is
higher than the energy splitting between the symmetric ground state
of the single exciton sector, and the higher energy states within
that sector, a further enhancement mechanism is the symmetric
coupling of higher energy single exciton states in ring A with
excited states {\it with the same symmetry type} in sector B.  The
transfer rate due to such couplings between states of the same
symmetry exhibits the {\it same} $MN$-fold enhancement as the
couplings between the symmetrized ground states.   Whether or not
such interplay between excited single-exciton states plays a
significant role depends on the strength of the various couplings
compared with each other and compared with the ambient temperature.
The full effects of differing symmetry types in such complex quantum
systems lies outside of the scope of the current paper and will be
dealt with in a further work.

\section{Application to nano-engineered LH2 complexes}

Even without a detailed master equation treatment, we can still
apply the concept of a `speeded up' cooperative sector to
experimentally observed effects \cite{Escalante10} reported an
anomolously long diffusion length in engineered arrays of LH2
complexes.  Two systems were investigated, a two-dimensional crystal
of LH2 complexes, and an effectively one-dimensional nanofabricated
array of such complexes. In both systems diffusion lengths of up to
a micron were reported.

On the face of it, this diffusion length seems absurdly long: LH2
complexes are about 7 nanometers in diameter and 6.8 nanometers in
height. A diffusion length of a micron requires $O(10^{5})$ exciton
hopping steps over the course of 1 to 1.5 nanosecond lifetime with
an effective displacement of about 300 units away from its original
location. If the dynamics is described by diffusive hopping from
complex to complex, such a diffusion length would require hopping
times of 10 to 15 femtoseconds.  By contrast, detailed calculations
and experimental observation of naturally occurring arrangements of
LH2 complexes suggest a hopping time of around 5 picoseconds, a
difference of almost three orders of magnitude.

Cooperative quantum behavior suggests an alternative explanation for
these anomolous diffusion lengths. Each LH2 complex contains a ring
of $N=18$ bacterial chlorophylls with a resonant frequency for light
with a wavelength of 800 nm, and 9 chlorophylls with a resonant
frequency at 850 nm.  Thus, we may reasonably expect a significant
enhancement from cooperative coherent effects. The frequencies and
linewidths of the LH2 complexes in the crystals and nanofabricated
arrays are essentially the same as in native LH2. The exact effect
of cooperative behavior depends on the precise arrangement of LH2
complexes in each array, and is difficult to calculate.  From the
above analysis, however, it is reasonable to suppose that
cooperative behavior enhances the actual LH2 to LH2 hopping rate by
a factor $\alpha$, For example, taking the `native' value of $5$
picoseconds for the unhanced hopping time, $\alpha = 5$ represents
an enhanced hopping time of around $1$ picosecond.

Even if $\alpha = 10$, we are still a factor a hundred away from the
10-15 femtosecond hopping rate required for an incoherent hopping
process to explain the anomolous diffusion rate.  Now, however, the
speeded up cooperative hopping allows a second effect of quantum
coherence to come into play. Let the lifetime of the exciton be $T$
and the decoherence time for the hopping process be $\tau$. We do
not know the decoherence time for hopping in such engineered arrays,
but in other photosynthetic systems it ranges from picoseconds to
tens of picoseconds. Call the hopping rate in the absence of
cooperative effects $\gamma$, and the hopping rate in the presence
of cooperative effects is $\alpha \gamma$. If $\alpha\gamma$ is
significantly greater than $1/\tau$, then the exciton hops {\it
coherently} through the array for about $ \ell  = \alpha \gamma
\tau$ steps.  In other words, for brief periods of time, the exciton
is performing a quantum walk.  During this time, as is usual with
quantum walks, the coherent diffusion time goes linearly in the
number of steps, rather than as the square root of the number of
steps. For times longer than the decoherence time, the hopping
becomes incoherent.  We can therefore model the combination of
coherent and incoherent hopping as diffusive transport with a
hopping rate equal to the original, unenhanced rate $\gamma$, but
with an increased step size of $\ell$. We see that the effect of
coherent hopping to increase the effective diffusion length.

Let the net number of units that the exciton has to diffuse before
decaying be $L$.  In the case of the LH2 complexes, $L$ is on the
order of 300. The effect of cooperative coherent behavior is to
reduce the total number of required incoherent hopping events from
$L^2$ to $L^2/\ell^2 = L^2/(\alpha\gamma\tau)^2$.  The total number
of incoherent hopping events in the coherence-enhanced diffusion is
no greater than the overall lifetime times the hopping rate,
 $T\gamma$.  Putting these relations together, to explain the anomolously long
diffusion rate in terms of cooperative coherent behavior, we require
that the dimensionless step size $\ell$ be at least
\begin{equation}
\ell = \alpha \tau \gamma > {L\over \gamma^{1/2} T^{1/2} }. \tag{7}
\end{equation}
For $L = 300$, $T=1$ nanosecond, $\gamma^{-1} = 5$ picoseconds, this
requires that $\alpha \tau$ be greater than $100$ picoseconds. For
example, if the coherent enhancement factor $\alpha = 5$, a
reasonable number given that there are 18 LH2 chlorophyll
participating in transport at 800 nm, and 9 participating at 850 nm,
then the hopping decoherence time must be at least 20 picoseconds.
If, by contrast, in the packed arrays of crystalline and
nanofabricated LH2, $\gamma=2$ picoseconds, then $\alpha\tau$ need
only be around $20$ picoseconds.    Pending more exact observations
of the hopping decoherence rate in LH2 arrays, and more exact
calculations of the actual coherent enhancement, such numbers seem
entirely reasonable: only moderate coherent enhancements are
required to explain the three orders of magnitude increase of the
apparent hopping rate in nano-engineered LH2 arrays.

\section{conclusion}

This paper presented a simplified discussion of the effects of
symmetry on hopping processes and environmental interactions.  Just
as symmetry provides an enhancement of superradiance, it can, under
the proper circumstances, lead to supertransport; i.e., an effective
enhancement of hopping rates and of coherent and incoherent
interactions with environmental modes. Conversely, antisymmetry can
significantly reduce effective environmental interactions and
enhance excitonic lifetimes. The interplay between symmetry and
antisymmetry properties of delocalized excitonic states in
photosynthetic complexes can be thought of as nature's use of
quantum coherence effects.

Full dynamical analysis of the effects of the symmetry breaking part
of the Hamiltonian will be the subject of a subsequent work
\cite{Mohseni10}. Silbey and co-workers constructed a theory for
multichromophoric Forster energy transfer that can provide a useful
model for studying enhanced cooperative transport in disordered
materials and photosynthetic complexes assuming that coupling
between collective donor and accepter chromophores can be treated
perturbatively \cite{JangSilbey04}. Generally, in addition to
important effects of disorders and phonon-bath couplings, the
geometrical structure of donor and acceptor in molecular aggregates
could play a significant role when one considers excitonic transfer
beyond dipole approximation \cite{Scholes02}. Recently there have
been a great deal of interest to study excitonic transport in
non-perturbative and non-Markovian regime \cite{Jang08,Ishizaki09}.
Similar approaches for exploiting certain symmetry of chromophoric
structures and their coherence/decoherence dynamical interplay
\cite{Mohseni08,CaoSilbey09} could lead to design principles for
engineering artificial excitonic systems, such as quantum dot
structures and organic materials, for efficient excitation energy
absorbtion, emission, storage, and transport. The observation of
anomolously long diffusion lengths in LH2 arrays suggests that even
moderate degrees of coherent cooperation can significantly enhance
the performance of engineered excitonic systems.

\noindent{\it Acknowledgements:} The authors acknowledge the support
of the W.M. Keck foundation, NSERC, Jeffrey Epstein, ENI, and
Lockheed Martin.

\end{document}